# Muon spin rotation studies of spin dynamics at avoided level crossings in LiY$_{0.998}$Ho$_{0.002}$F$_4$

M. J. Graf [1*], J. Lago [2], A. Lascialfari [2,3], A. Amato [4], C. Baines [4], S. R.Giblin [5], J. S. Lord [5], A. M. Tkachuk [6], and B. Barbara [7]

[1] Department of Physics, Boston College, Chestnut Hill, MA 02467 USA
[2] Department of Physics "A. Volta", University of Pavia, and CNR-INFM, I27100 Pavia, Italy
[3] Inst. of General Physiology and Biological Chemistry, University of Milano, I20134 Milano, Italy
[4] Paul Scherrer Institute, CH 5232 Villigen PSI, Switzerland
[5] Rutherford Appleton Laboratory, Chilton Didcot, Oxfordshire OX11 0QX, UK
[6] St. Petersburg State University of Information Technology, Mechanics and Optics, 199034, 4, Birzhevaja line, St. Petersburg, Russia
[7] Institut Néel, Département Nanosciences, CNRS, 38042 Grenoble Cedex-09, France

We have studied the Ho$^{3+}$ spin dynamics for LiY$_{0.998}$Ho$_{0.002}$F$_4$ through the positive muon (μ$^+$) transverse field depolarization rate $\lambda_{TF}$ as a function of temperature and magnetic field. We find sharp reductions in $\lambda_{TF}(H)$ at fields of 23, 46 and 69 mT, for which the Ho$^{3+}$ ion system has field-induced (avoided) level crossings. The reduction scales with calculated level repulsions, suggesting that μ$^+$ depolarization by slow fluctuations of non-resonant Ho$^{3+}$ spin states is partially suppressed when resonant tunneling opens new fluctuation channels at frequencies much greater than the muon precession frequency.

Zero-dimensional magnetic systems exhibit spin dynamics which are dominated by quantum effects at low temperatures. This has been made evident by recent studies of single-molecule magnets (SMM) [1]. In these systems a cluster of transition metal ions is imbedded within a large organic molecule, with the ions becoming locked into either non-zero (e.g., Mn12 or Fe8) or zero-spin states (Cr8) when the thermal energy $k_B T$ is much smaller than the exchange coupling. The easy-axis anisotropy produces a large energy barrier (typically several tens of degrees K) to spin reversal, which can be overcome by thermal activation, quantum tunneling, or a combination of both mechanisms. Application of a magnetic field can produce level-crossings through Zeeman splitting of the low lying energy states, and mixing of the states through either intrinsic (e.g., transverse crystal-field and hyperfine fields, disorder) or extrinsic (applied transverse fields) processes produces tunnel splittings and avoided-level crossings (ALCs). Enhanced low-temperature spin transition rates due to state mixing are observed at magnetic fields corresponding to ALCs and produce abrupt changes in the sample magnetization, as first observed in Mn12 [2].

NMR is an effective probe for studying the spin dynamics in these systems in moderate-to-strong magnetic fields [1]. At low temperatures ($T \ll \Delta$, where $\Delta$ is the gap between the ground state and the first excited state) in selected high-symmetry systems (e.g. molecular rings) the enhanced transition rate at ALCs results in peaks in the magnetic field dependence of the spin-lattice relaxation rate $1/T_1$. One might expect positive muon spin rotation/relaxation ($\mu^+$SR) to be an important complementary probe to the NMR measurements on SMM, as one has the possibility of studying the spin dynamics with one technique from zero to high magnetic fields; moreover the local field can be probed at locations other than the active nuclear site. In fact, although $\mu^+$SR is an extremely sensitive local magnetic probe, we know of no example where measurements have exhibited an enhanced spin transition rates at ALCs in SMM systems. In part this can be attributed to the difficulty in producing large single crystals, resulting in most studies being conducted on powdered samples [3].

The *single-ion magnet* $LiY_{1-x}Ho_xF_4$ exhibits many of the same properties as SMMs. The $J=8$ electronic levels of the $Ho^{3+}$ ion are split by the crystal electric field producing a low-lying ground state doublet, and hyperfine coupling with the $^{165}Ho$ nucleus ($I=7/2$) further splits the ground state doublet. An external magnetic field component along the quantization axis (c-axis in this Scheelite structure) produces ALCs for $H_n = 23\,mT$, with $-7 \leq n \leq 7$ and $n$ being an integer (see Fig. 1). Thus for small $x$, where the Ho-Ho dipolar coupling is weak, the spin dynamics in this system are similar to those for the SMMs, e.g., magnetization measurements show steps at the ALCs at low temperatures [4]. Recent $^{19}F$ NMR low temperature measurements of the magnetic field dependence of $1/T_1$ exhibit strikingly sharp peaks at magnetic field values corresponding to ALCs for $n = 4-7$ [5].

Detailed calculations [4] of the variation of the single-ion electronuclear energy levels with magnetic field along the quantization axis predict both the field locations of the ALCs and the degree of state mixing in the absence of crystalline defects or a component of the applied magnetic field perpendicular to the c-axis. The resultant tunnel splittings arise from the combined effect of the crystal field and the hyperfine interactions. The crystal-field acting on the total angular momentum ***J*** gives a quasi-Ising ground state doublet $\Gamma_{3,4}$, with the next highest levels being singlets, resulting from the same $\Gamma_2$, at 10 K and 36 K above the ground-state, and two nearly-degenerate $\Gamma_1$ states at 70 K above the ground state. The hyperfine interactions given by the Hamiltonian

$$H = A_J \left[ J_z I_z + (1/2)(J_+ I_- + J_- I_+) \right], \quad (1)$$

where $A_J$ is the hyperfine constant and $J_i$ and $I_i$ are the electronic and nuclear angular momentum components, respectively, multiply the crystal-field degeneracy of each state by $2I+1=8$. First order perturbation calculations modify the crystal-field energy by $A_J(M \pm m_I)$, where $M$ and $m_I$ are the electronic and nuclear spin quantum numbers, showing that each doublet (in particular $\Gamma_{3,4}$) transforms into a comb of magnetic field dependent levels with 8 levels going up (effective electronic spin 1/2) and 8 levels going down (effective electronic spin -1/2). Including the off-diagonal part of Eq. (1), second order perturbation theory shows that the states of $\Gamma_{3,4}$ mix with those of $\Gamma_2$ or $\Gamma_1$ when the selection rule $\Delta m_I/2 = \pm 1$ is fulfilled. The resultant repulsion at the crossings of the $\Gamma_{3,4}$ states increases with the repulsion of singlets. This is why, in Fig.1, the crossings at $n = \pm 1, \pm 3, \text{and} \pm 5$ and energy -179.55 K, have large tunnel splittings of 25 mK, 22 mK and 16 mK, respectively (mixing via $\Gamma_2$), while all other splittings are very small. This energy level diagram has been confirmed in detail by recent EPR measurements [6].

In this work we report transverse field µ$^+$SR measurements on LiY$_{0.998}$Ho$_{0.002}$F$_4$. This system offers many distinct advantages over SMM for µ$^+$SR studies, the most obvious being that large, high quality and well-characterized single crystals can be studied. Earlier longitudinal field µ$^+$SR measurements [7] showed a very weak increase in the muon depolarization rate at the $n=1$ ALC, but the quantizing field of 23 mT significantly depressed the muon depolarization rate, making it extremely difficult to undertake any detailed quantitative studies. In this work we show that transverse field µ$^+$SR is a highly sensitive probe of the Ho$^{3+}$ spin dynamics, and we find sharp *decreases* in the muon depolarization rate at fields corresponding to avoided level crossings, with relative sizes in good agreement with the calculations [4] described above.

A large single crystal with nominal Ho concentration $x = 0.002$ was grown and cut into plates with dimensions 3x8x30 mm$^3$, with the crystalline c-axis aligned along the 3mm edge. We measured the $^{19}$F $1/T_1$ as a function of magnetic field at 1.7 K for one of these plates and confirmed the presence of sharp peaks in $1/T_1$ at ALCs with $n = 5, 6,$ and $7$, as described for a different crystal in Ref. 5.

In time-differential muon spectroscopy, a spin polarized positive muon enters a sample where it quickly (10 *ns* or less) comes to rest. At some later time *t* the muon decays, and the resulting positron is preferentially emitted along the muon spin polarization axis and detected. After several million independent decay events one constructs a time histogram of the detector signal asymmetry $A(t)$ [8] which is proportional to the muon depolarization function $P_\mu(t)$ in the local field at the muon stopping site. $P_\mu(t)$ is defined as the projection of the spin polarization along the initial polarization direction. In the transverse mode, the initial muon polarization is perpendicular to the applied magnetic field, resulting in muon precession. A variation of the local magnetic field between muon stopping sites will produce a dephasing of $P_\mu(t)$ due to the spread of Larmor frequencies for the individual muon decay events. Thus our measurements are extremely sensitive to magnetic inhomogeneity. The μ$^+$SR measurements were conducted utilizing the GPS and LTF spectrometers on the πM3 continuous beamline at the Paul Scherrer Institute (PSI). The beam momentum and external magnetic field were aligned parallel to the crystalline c-axis. Experiments were conducted in a gas flow cryostat and a dilution refrigerator so that our measurements span the range $30\,\text{mK} \leq T \leq 50\,\text{K}$. We utilized the spin-rotated mode, where the muon spin is aligned at an inclination of approximately 50 degrees with respect to the beam momentum. The transverse depolarization is then monitored with detectors oriented along a line perpendicular to the beam momentum.

Earlier zero field $\mu^+$SR results on these samples [7] showed that nearly all the muons are trapped by pairs of adjacent F$^-$ ions, forming the stable three-spin system F-$\mu$-F [9]. In zero-magnetic field this system produces spin-flip oscillations at three non-zero frequencies $(3-\sqrt{3})\omega_D/2$, $\sqrt{3}\,\omega_D$, $(3+\sqrt{3})\omega_D/2$, where $f_D = \omega_D/2\pi$ is determined by the average separation between the muon and the F ions and their gyromagnetic ratios. The relative amplitudes of these components are determined by the orientation of the muon polarization relative to the line joining the adjoining F ions. Our analysis of the zero-field data indicates that: (1) the most likely muon stopping sites are the four magnetically equivalent locations in the unit cell where two F ions are closest, and the F ions are pulled in from a separation of 0.260 nm to 0.239 nm by the muon; (2) below 10 K the damping of the oscillations increases significantly; and (3) the fitted value of $f_D$ increases slightly below 6 K, from 0.211 MHz to 0.220 MHz. The first result is consistent with results on the F-$\mu$-F formation in comparable insulating fluorides. The second result is consistent with the transverse field results presented and discussed in detail below, which show the onset at low temperatures of quasi-static magnetic disorder associated with the Ho$^{3+}$ ions. Monte Carlo calculations for F-$\mu$-F oscillations in the presence of a Gaussian distribution of static moments of width $\Delta$ show that the observed $f_D$ in zero field will have a small positive correction that increases as $\Delta^2$, which provides a qualitative understanding of the third result above [10].

We now describe the variation of the depolarization rate with temperature in a magnetic field of 23 mT, corresponding to the $n=1$ series of avoided level crossings (Fig. 1). The oscillatory asymmetry $A(t)$ exhibits a damping which can be readily fit by a stretched exponential function, so that $A(t) = A_0 \exp\left[-(\lambda_{TF}\, t)^\beta\right]\cos(\omega_L t + \phi)$, where $\omega_L$ is the Larmor frequency. Representative fits are shown in Fig. 2a for temperatures of 50 K and 1.8 K. All $\chi^2$ values (per degree of freedom) lie in the range 0.9997 – 1.0313. In Fig. 2b we show the temperature dependence of the depolarization rate $\lambda_{TF}(T)$; the inset shows the variation of the exponent $\beta$ with temperature. At $T = 50\,\text{K}$ $\lambda_{TF}$ has a relatively large value of $0.409 \pm 0.003$ µs$^{-1}$, reflecting the quasi-static magnetic disorder produced by the various magnetic nuclear species. As the temperature is lowered it remains roughly constant until about 20 K, below which it increases sharply, reaching a value of $\lambda_{TF} = 0.866 \pm 0.008$ µs$^{-1}$ at 30 mK. The increase of $\lambda_{TF}$ below 20 K results from decoherence of the precessional motion of muon spins due to the increase of quasi-static disorder at temperatures lower than the barrier for spin reversal. The $\Gamma_2$ singlets, which are roughly 10 K and 36 K above the ground-state, induce strong longitudinal and transverse fluctuations for $T > 10\,\text{K}$, which defines the top of the spin-reversal energy barrier. These spin fluctuations are too fast to affect the local disorder on the muon precessional timescale. We confirmed that the high temperature depolarization is caused primarily by the magnetic nuclei by observing that the depolarization at $T = 50\,\text{K}$ for samples with higher Ho concentrations ($0.005 \leq x \leq 0.1$) are identical to the results for $x = 0.002$. Upon cooling below the barrier of 10 K, the spin-phonon relaxation times become longer (74 µs at 5 K and 1.5 ms at 2 K) and the Ho$^{3+}$ spin disorder becomes nearly static relative to the muon precession times. We see no sign of saturation of $\lambda_{TF}$ down to the lowest temperature studied (30 mK). Assuming that the transverse field depolarization is dominated by quasi-static disorder of electronic and nuclear spin, we infer that there may still be a small dynamic component to the local field at low temperatures, as observed in longitudinal field studies on $\text{LiY}_{0.955}\text{Ho}_{0.045}\text{F}_4$ at very low temperatures [11]. The increase in $\lambda_{TF}$ at low temperatures becomes larger with increasing $x$, confirming that the depolarization is related to the magnetism associated with the Ho ion. The concentration dependence of the zero, transverse, and longitudinal field depolarization will be described in a future publication.

The onset of quasi-static spin disorder of the $\text{Ho}^{3+}$ ions below 20 K is also evident in the temperature dependence of the exponent $\beta$. A concentrated system of randomly oriented nuclear dipoles is expected to have an exponent of 2, while for a dilute system of paramagnetic spins, the exponent is 1 for quasi-static moments, and 0.5 for fluctuating moments [12]. The observed exponent is roughly 1.5 at high temperatures ($T > 20\,\text{K}$), reaches a minimum of 1.1 near 5 K, and then increases to a value near 1.2 at low temperatures. The minimum in the exponent at 5 K indicates that the fluctuation rate is now comparable to the muon precession rate. At lower temperatures, the Ho fluctuations become quasi-static and the exponent increases once again.

To probe the spin dynamics in the vicinity of the ALCs, we measured the magnetic field dependence of $\lambda_{TF}$ at temperatures $T \leq 3\,\text{K}$ over the range $18\,\text{mT} \leq H \leq 80\,\text{mT}$. We once again fit the depolarization to a stretched exponential function, with the exponent increasing slightly from 1.16 ± 0.02 at 18 mT to 1.26 ± 0.02 at 80 mT. Data taken at 1.8 K are shown in Fig. 3. The remarkable feature of the curve is the presence of sharp minima in $\lambda_{TF}$ at 23 mT, 46 mT, and 69 mT, the field values at which the first three ALCs occur. Moreover, it is readily seen that the reductions are quite large for $n = 1$ and $3$, while the reduction for $n = 2$ is barely discernible. As discussed below, these minima are produced by the onset of new channels for $\text{Ho}^{3+}$ spin fluctuations which are much faster than the muon precession rates (< 10 MHz), thereby reducing the (time-averaged) local magnetic disorder and decreasing $\lambda_{TF}$.

In general, below 4 K the spin disorder associated with the $Ho^{3+}$ dipolar field is primarily quasi-static. However, at the resonance fields $H_n = 23\,mT$, energy levels are nearly degenerate, and the $Ho^{3+}$ will tunnel between the coupled energy levels with a frequency determined by the tunnel splitting at the ALC. This leads to a natural interpretation of the data presented in Fig. 3. At the ALCs we have a new channel for enhanced spin fluctuation rates at low temperatures due to mixing between the Ho hyperfine-split states. At $n=1$ and $3$ the largest tunnel splittings are predicted to be roughly 25 mK, corresponding to a characteristic frequency of 400 MHz. These fluctuations, about 100 times faster than muon precession times, will then effectively remove a significant portion of the spin disorder, producing a smaller spread in the time-average local field at the muon site and a decrease of $\lambda_{TF}$. At $n=2$ the predicted splitting, and therefore the tunneling frequency, is very much smaller and so the reduction in $\lambda_{TF}$ is much less prominent.

The temperature dependence of $\lambda_{TF}$ in the vicinity of the $n=1$ ALC (Fig. 4) is in qualitative agreement with the above mechanism. The sharp minimum at 23 mT is approximately constant in size and width for temperatures between 3 K and 0.9 K. However, at 0.5 K the minimum becomes weaker, and is no longer present at 30 mK. From Fig. 1 we see that at 23 mT ($n=1$), the ALC with strong mixing lies roughly 600 mK above the lowest energy level. Fast resonant tunneling will occur for those $Ho^{3+}$ ions which populate these two (nearly) crossing levels. Below 600 mK the thermal population of these states will be reduced, and so the sharp minimum in $\lambda_{TF}$ is expected to disappear, as observed.

We now confirm that such a low concentration of Ho ions can significantly alter the muon spin dynamics. The dipolar magnetic field contributions at the muon site from a nearby $^{19}F$ nucleus and Ho$^{3+}$ ion vary as $\left(\mu_i/r_i^3\right)$, where $\mu_i$ and $r_i$ are the magnetic moment and distance from the muon for the $i^{th}$ magnetic species, respectively. The muon is located at 0.115 nm from two $^{19}$F nuclei, while a Ho$^{3+}$ ion has a magnetic moment which is about 5000 times larger than that for the $^{19}$F, so the Ho$^{3+}$ would produce a comparable dipolar field at the muon site to that of the $^{19}$F nucleus if located roughly 2 nm from the muon. For $x=0.002$, the average spherical volume per Ho ion has a radius of 2.3 nm [13], and so it is quite reasonable to expect that the local field contributions of the Ho$^{3+}$ ions and $^{19}$F nuclei will be comparable.

Finally, we comment on the possible influence of the F-μ-F nuclear dipolar interactions on our observations. In transverse field the isolated F-μ-F system would show a central line at the muon Larmor frequency and two sidebands with splitting of order $f_D$ (angle-dependent). These are just observable in the Fourier transform of high temperature data, but become indistinguishable from the central peak at low temperatures where broadening is increased. Monte-Carlo simulations [10] show that in transverse fields the effect of additional interactions (electronic or nuclear) is to broaden all three lines equally, and the effective $f_D$ does not vary significantly. The fitted relaxation function covers all three lines, so $\lambda_{TF}$ will have a constant contribution from the F-μ-F splitting. We note that at 1.8 K the dc magnetization is a smooth function of magnetic field, and so no abrupt changes in the static magnetic environment occur at ALCs. We conclude that our fundamental result, that the change in $\lambda_{TF}$ at ALCs occurs due to a change in the spin dynamics of the Ho$^{3+}$, remains unchanged. Nonetheless, it would be interesting to determine the exact role of F-μ-F oscillations by repeating these measurements with the muon decoupled from the fluorine ions, e.g., by applying an appropriately tuned RF field.

It is remarkable that a simple electro-nuclear crystal field calculation allows us to accurately predict not only the location of level crossings, but also the degree of state mixing. This is possible because of the high crystalline quality and the excellent knowledge of the crystal field and hyperfine parameters. In this sense the rare-earth single-ion systems have a great advantage over single-molecule magnet systems, for which the local magnetic environment can be very complicated and can have a fairly high degree of disorder. This allows for the precise characterization, and ultimately tailoring, of the magnetic properties for applications in, for example, quantum computing or information storage. Indeed, a related rare-earth-based system, $Er^{3+}$:$CaWO_4$, has already been demonstrated [14] to be suitable for use as a quantum q-bit.

In conclusion, we have demonstrated that transverse field $\mu^+$SR is a highly effective probe of the spin dynamics at avoided level crossings of the $Ho^{3+}$ ion doped into $LiYF_4$. Below 20 K the muon depolarization rate increases rapidly at the onset of quasi-static spin disorder associated with long relaxation times spin-phonon relaxation times of the $Ho^{3+}$ ion. However, at avoided level crossings the rapid depolarization due to disorder can be circumvented by the fast $Ho^{3+}$ fluctuations resulting from quantum mixing of the states. The relative sizes and temperature dependence of the observed reductions in depolarization rates agree with the degree of level mixing predicted by perturbation theory based on a coupling of the ground state and excited electronic levels via transverse hyperfine interactions.

The authors would like to thank Pietro Carretta, Ferdinando Borsa, Robert Kiefl, and members of the Avoided Level Crossing Group at PSI for helpful discussions, and Jeffrey Klatsky for assistance with some of the data analysis. This work was supported by grants from the Petroleum Research Fund of the American Chemical Society, National Science Foundation DMR-0710525, INTAS 03-51-4953, and the QUEMOLNA and MAGMANET European-NMP programs. Experiments were performed at the Swiss Muon Source, Paul Scherrer Institute, Villigen, Switzerland.

**Figure Captions**

**Figure 1**. Calculations of the magnetic field variation of the hyperfine split ground state electronic energy levels. Inset: closer view of the dominant tunnel splittings at an energy of approximately -179.55 K (adapted from Ref. 6).

**Figure 2**. (a) Time-dependent asymmetry at two temperatures. The curves are offset for clarity. The solid lines are stretched exponential fits, which have $\chi^2$ values per degree of freedom of 1.026 (50 K) and 1.018 (1.8 K). (b) The temperature variation of the depolarization rate at a fixed field of 23 mT. Inset: Temperature dependence of the exponent.

**Figure 3**. Magnetic field variation of transverse field depolarization rate versus applied field at $T = 1.8\,\text{K}$.

**Figure 4**. Magnetic field variation of the transverse field depolarization rate $\lambda_{TF}$ in the vicinity of the $n = 1$ avoided level crossings at several temperatures. The curves are offset for clarity.

**Figure 1**

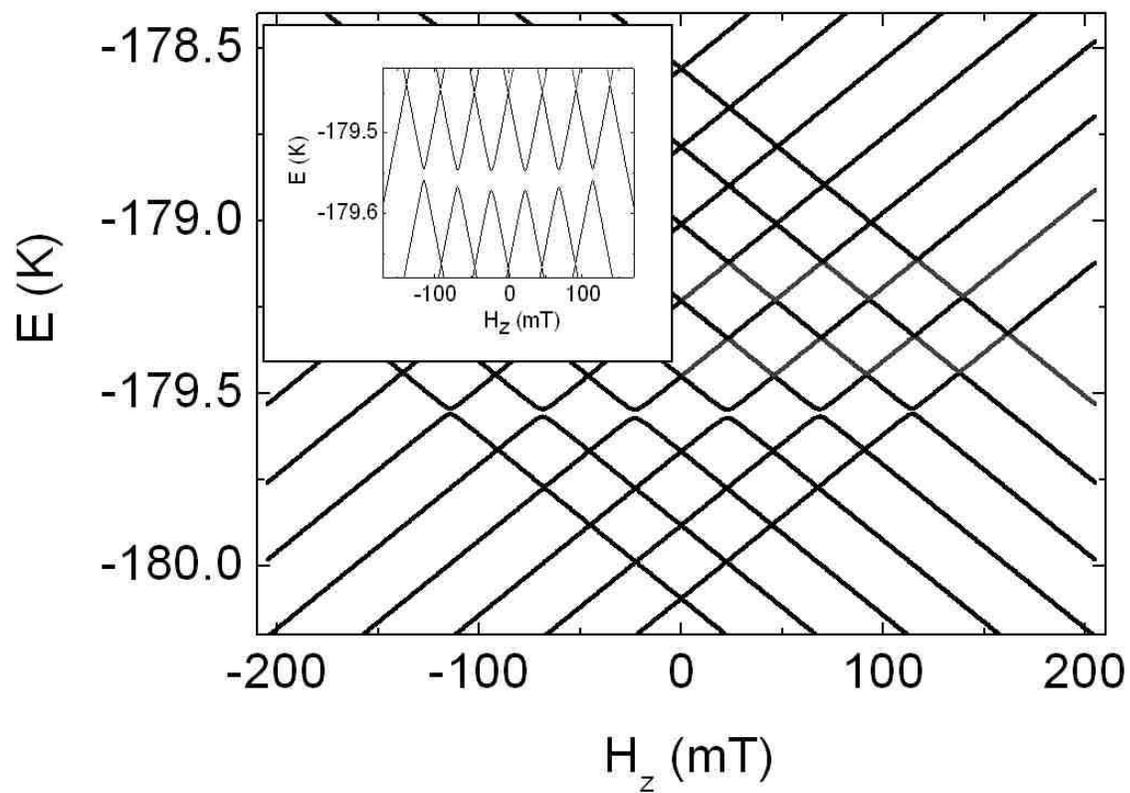

**Figure 2**

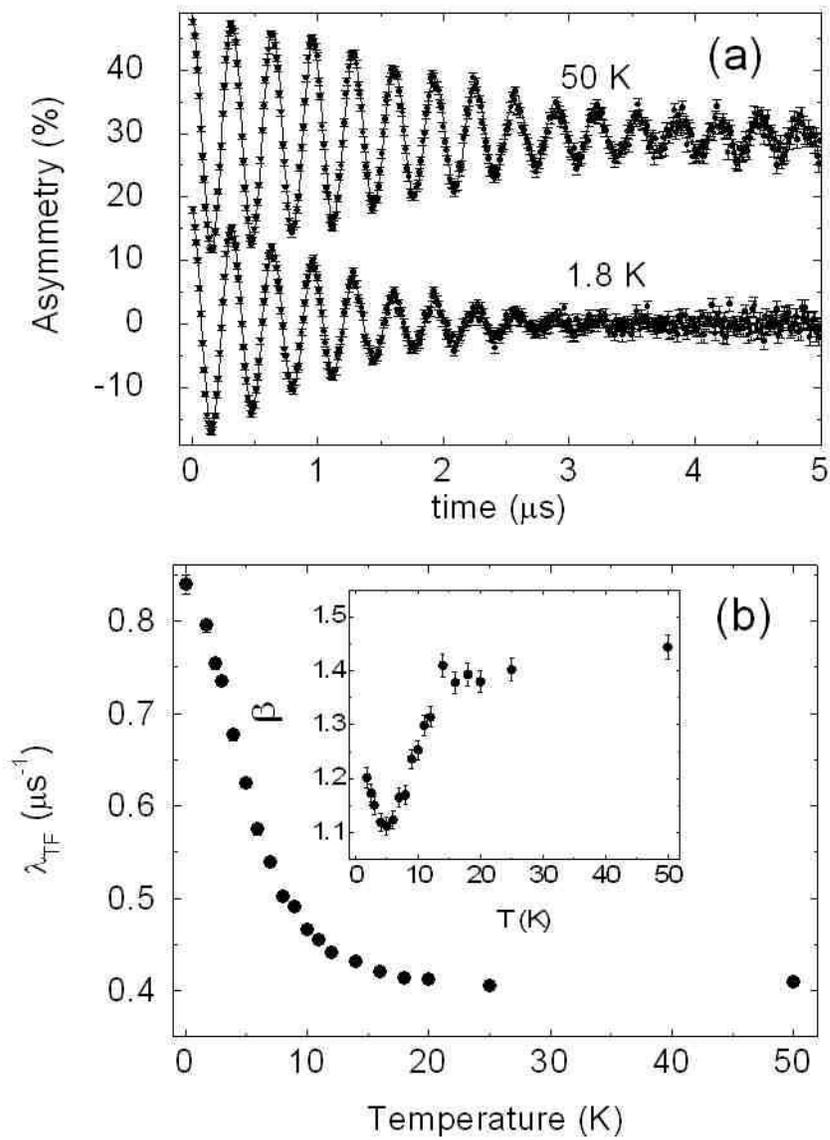

**Figure 3**

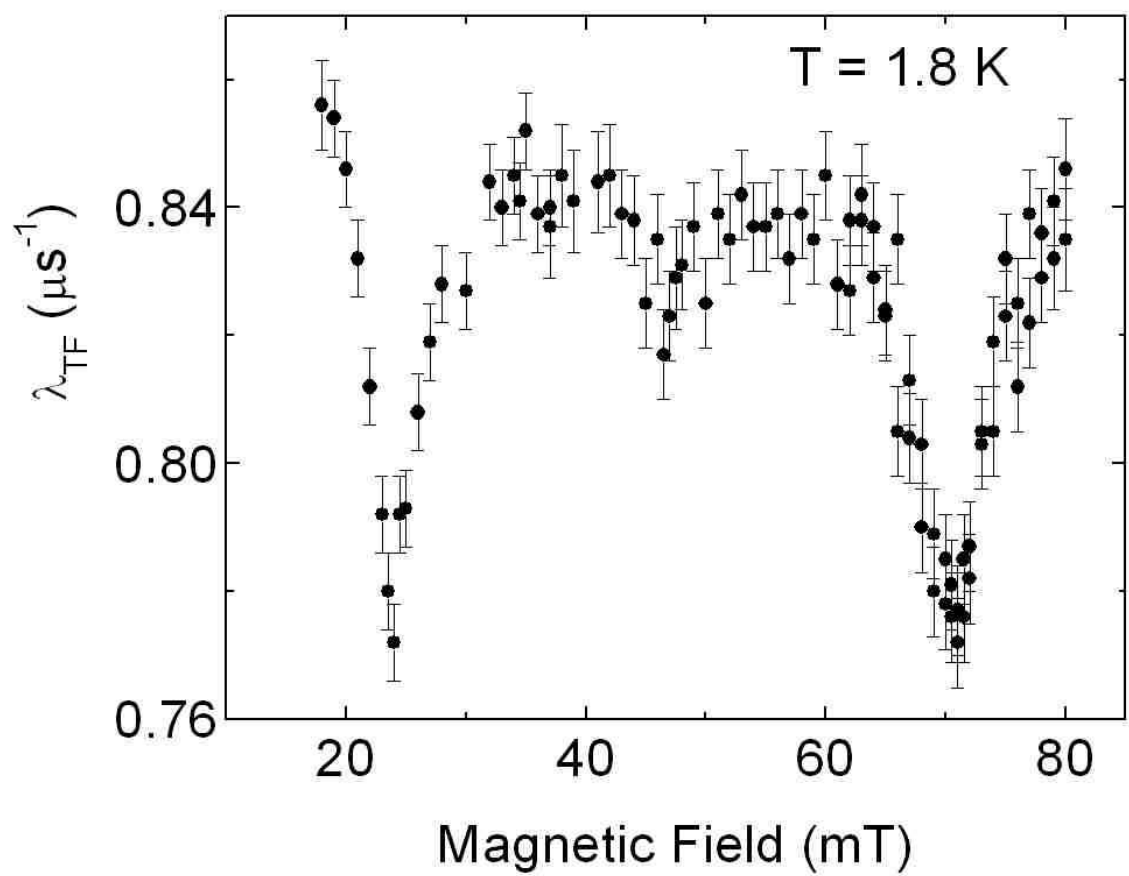

**Figure 4**

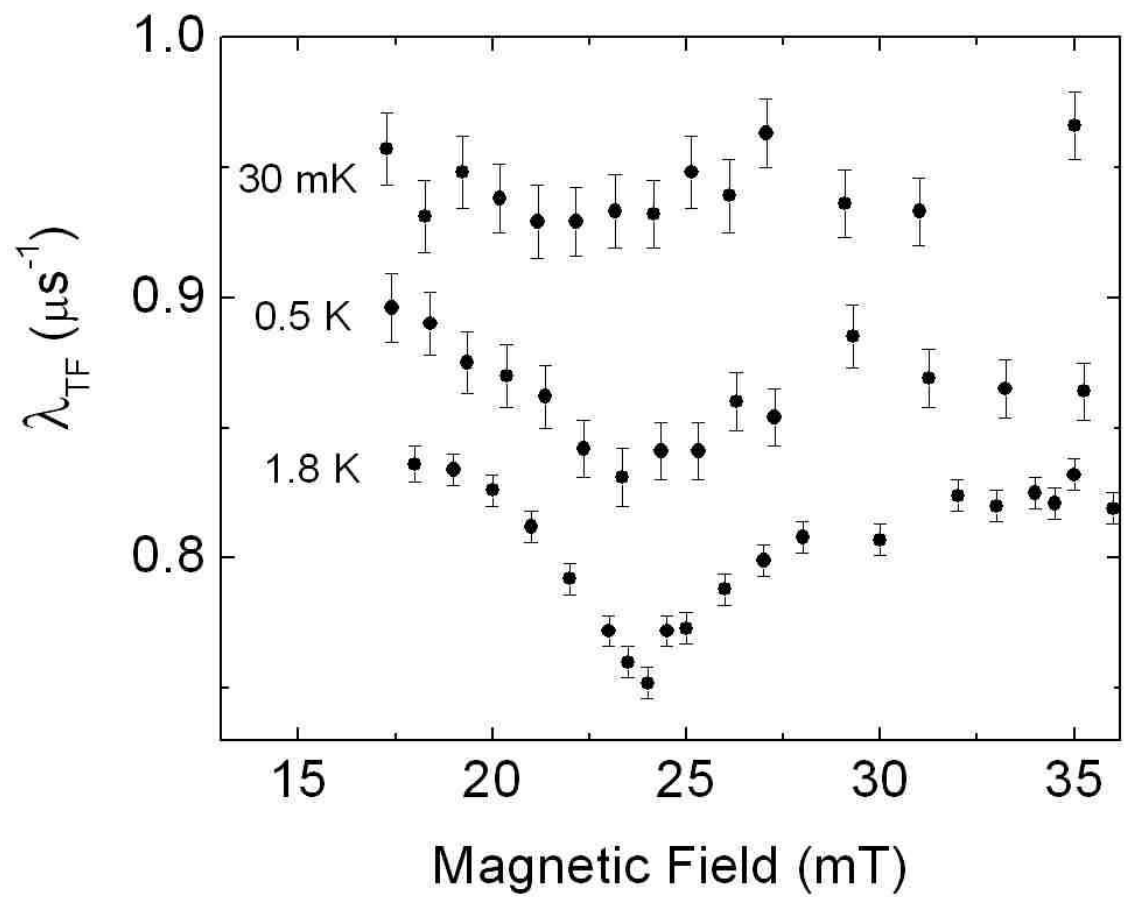